\documentclass{emulateapj}

\def\lax{{$\mathrel{\hbox{\rlap{\hbox{\lower4pt\hbox{$\sim$}}}\hbox{$<$}}}$}}
\def\gax{{$\mathrel{\hbox{\rlap{\hbox{\lower4pt\hbox{$\sim$}}}\hbox{$>$}}}$}}

\slugcomment{}

\shorttitle{A new approach for probing circumbinary disks}
\shortauthors{Hayasaki \& Okazaki}

\begin{document}


\title{A new approach for probing circumbinary disks}


\author{Kimitake Hayasaki\altaffilmark{1}}
\affil{Yukawa Institute for Theoretical Physics, Kyoto University, Oiwake-cho,
Kitashirakawa, Sakyo-ku, Kyoto 606-8502, Japan}
\email{kimitake@yukawa.kyoto-u.ac.jp}
\and
\author{Atsuo T. Okazaki\altaffilmark{2}}
\affil{Faculty of Engineering, Hokkai-Gakuen University, Toyohira-ku, Sapporo
062-8605, Japan}


\begin{abstract}
Circumbinary disks are considered to exist in a wide variety of
astrophysical objects, e.g., young binary stars, protoplanetary
systems, and massive binary black hole systems in active galactic
nuclei (AGNs). 
However, there is no definite evidence for the circumbinary disk 
except for some in a few young binary star systems. 
In this Letter, we study possible oscillation modes 
in circumbinary disks around eccentric and circular binaries.
We find that progarde, nonaxisymmetric waves are
induced in the inner part of the circumbinary disk by the tidal potential of the binary.
Such waves would cause variabilities in emission line
profiles from circumbinary disks. Because of prograde precession
of the waves, the distance between each component of the binary and
the inner edge of the circumbinary disk varies with the beat period
between the precession period of the wave and the binary orbital
period. As a result, light curves from the circumbinary disks are
also expected to vary with the same period. The current study thus
provides a new method to detect circumbinary disks in various
astrophysical systems.
\end{abstract}


\section{Introduction}
\label{sec:intro}

Astrophysical disks are ubiquitous 
in the various systems of the universe: 
star-compact object systems, 
star-planet systems, 
active galactic nuclei (AGNs), and so forth.
These disks surround the individual objects as a 
circumobject disk. If the object surrounded by a rotating disk is a binary, the disk is
called a circumbinary disk 
(see Figure~\ref{fig:system} for a schematic view of the circumbinary disk).

About 60$\%$ of main sequence stars are considered to be born
as { binary or multiple systems \citep{dm91}. 
Numerical simulations have confirmed that 
young binary stars embedded in dense molecular gas 
have a circumbinary disks \citep{al96a, bb97, gk1, gk2} 
and that a circumsteller disk is also formed 
around each star \citep{bb97,gk1,gk2}.
Indeed, direct imaging of the circumbinary disk was 
successfully achieved by 
interferometric observations of a few young binary systems
including GG Tau \citep{dut94} and UY Aur \citep{duv98}.

In the early stage of planet formation, 
a planet orbiting a star will be embedded in a rotating disk 
(hereafter, circumbinary disk) surrounding them. 
\cite{kd06} showed, by performing numerical simulations, 
that the cicrumbinary disk becomes eccentric 
due to the resonant interaction between the planet and the disk.
Such a planet--disk interaction also causes the 
significant evolution of the orbital elements of the planet, 
which gives a possible explanation about the observed 
high orbital eccentricities in extrasolar planetary systems \citep{gs03}.
The direct probing of the circumbinary disk is, therefore, a key to testing this scenario.

Massive black holes are considered to 
co-evolve with their host galaxies \citep{mag98,fm00,geb00}.
There is inevitably an evolutionary stage as a binary black hole in the course of
galaxy merger until the coalescence of two black holes \citep{bege80,may07,kh08}. 
\cite{haya07} found that if a binary black hole is surrounded by a circumbinary disk, 
mass is transferred from the disk to each black hole. 
The system then has a triple disk composed of 
an accretion disk around each black hole and a circumbinary disk 
as a mass reservoir around the binary \citep{haya08}.
There is, however, still little evidence for the circumbinary disk as well as for
the binary black hole itself. 
Therefore, the detection of the circumbinary disk is 
also one of the important scientific motivations in probing massive binary black holes
with parsec/subparsec separations.
Quite recently, \cite{bt08} and \cite{do08} proposed the hypothesis 
that SDSSJO92712.65+294344.0 is a massive binary black hole, 
by interpreting the observed emission line features as those arising 
from the mass-transfer stream from the circumbinary disk.

There are many phenomena caused by oscillations 
in circumstellar/accretion disks. 
One of the most famous among them is superhumps.
A superhump is a periodic luminosity hump 
on the light curves in an accreting binary system, 
with a slightly longer period than the orbital period of the binary.
The superhump phenomenon was first discovered in the SU Ursae Majoris
class of dwarf novae, which consists of a white dwarf and a late-type star 
with a low mass ratio, less than 0.2 \citep{pat79,vogt80}.
It is attributed to the precession of a deformed accretion disk induced 
by the tidal potential of the binary \citep{osaki85}.
{\cite{lu91} showed that the deformation of the disk
is due to the growth of an eccentric (i.e., $m=1$) perturbation through nonlinear coupling
with the tidal potential.}
Hitherto, no phenomena caused by oscillation modes have been detected in
circumbinary disks.

In this Letter, we investigate the tidally induced oscillation modes 
in circumbinary disks. 
These modes can be driven by resonantly excited modes  at particular resonance radii, 
which are similar to the ones responsible for superhumps in dwarf novae systems. 
The Letter is organized as follows. 
In Section 2, we derive the azimuthally and temporally averaged 
tidal potential around an eccentric binary and discuss 
the possible oscillation modes and their frequencies in circumbinary disks. 
Section~3 is devoted to a summary and discussion.

\begin{figure}
\resizebox{\hsize}{!}{
\includegraphics*[width=86mm]{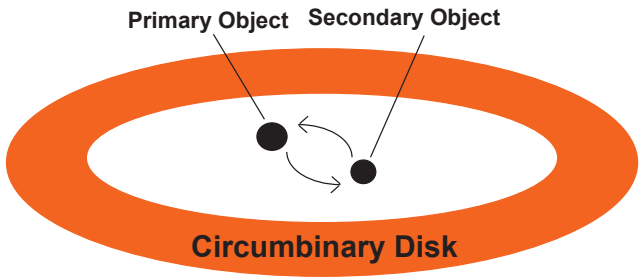}
}
 \caption{
 Schematic diagram of a circumbinary disk surrounding the primary object
 and the secondary object which are gravitationally bound as a binary.
 }
 \label{fig:system}
\end{figure}


\section{Tidally induced spiral waves}

We consider two gravitationally bound point masses 
surrounded by a circumbinary disk.
The point masses follow Kepler's third law, 
while the geometrically thin circumbinary disk 
has a rotation slightly deviated from Keplerian rotation
because of the tidal potential of the binary.
We assume that the circumbinary disk is truncated at an inner radius, $r_{\rm in}$,
by the resonant interaction with the binary (e.g., \citealt{al94}),
but extends outward to a large distance.

\subsection{Tidal potential}
{First}, we consider the tidal potential around an eccentric binary.
In the inertia{l} frame of reference centered on the common center of the binary,
the gravitational potential can be written as
\begin{eqnarray}
\Phi(r,\theta,t)
&=&-\frac{GM_{1}}{|\bf{r}-\bf{r}_{1}|}-\frac{GM_{2}}{|\bf{r}-\bf{r}_{2}|} \nonumber \\
\nonumber \\
&=&-\frac{GM_{1}}{\sqrt{r^{2} -2rr_{1}\cos (\theta-f)+r_{1}^2}} \nonumber \\
&&
-\frac{GM_{2}}{\sqrt{r^{2} -2rr_{2}\cos (\theta-f+\pi)+r_{2}^2} }
\label{eq:Phi}
\end{eqnarray}
where $M_{1}$ and $M_{2}$ are the masses of the primary 
and secondary components, respectively, 
$\bf{r}_{1}$ and $\bf{r}_{2}$ are their position vectors,
and $f$ is the true anomaly of the primary component.
Here, $r_{i}\hspace{1mm}(i=1, 2)$ periodically varies with time 
because of the orbital eccentricity and is given by
\begin{eqnarray}
\frac{r_{i}}{a}=\frac{\eta_{i}(1-e^2)}{1+e\cos{f}},
\label{pvi}
\end{eqnarray}
where $a$ is the semimajor axis and 
$\eta_{1}$ and $\eta_{2}$ are $q/(1+q)$ and $1/(1+q)$ with mass ratio 
$q=M_{2}/M_{1}$, respectively.

Azimuthally and temporally averaging equation~(\ref{eq:Phi}) 
with an assumption of $r\gg{r_{1}}$ and $r\gg{r_{2}}$, 
we obtain the averaged binary potential over one orbital period, $\bar{\Phi}(r)$, 
in the form:
\begin{eqnarray}
\bar{\Phi}(r)&=&\frac{1}{4\pi^2}\int_{0}^{2\pi}\int_{0}^{2\pi}\Phi(r,\theta,t)d\theta{d(\Omega_{\rm{orb}}t)}
\nonumber \\
&=&
-\frac{GM}{r}\left[1+\frac{1}{4}\left(\frac{a}{r}\right)^2\frac{q}{(1+q)^2}\left(1+\frac{3}{2}e^2\right)\right],
\label{eq:Phi_r}
\end{eqnarray}
where $M$ is the binary mass $M_{1}+M_{2}$. 
Here, we have neglected terms of the order of $e^{3}$ or higher and made use of
the following expansion formula: 
\begin{eqnarray}
\frac{r_{i}}{a}
&&
=\eta_{i}\left[
1-e\cos\Omega_{\rm{orb}}t + \frac{e^2}{2}(1-\cos2\Omega_{\rm{orb}}t) + O(e^3) 
\right]\nonumber
\label{eq:expansion}
\end{eqnarray} 
\citep[e.g.,][]{md99}. 
Note that these series converge for $e \lesssim 0.66$.

The second term on the right-hand side of equation~(\ref{eq:Phi_r})
gives the tidal perturbation potential
around the eccentric binary:
\begin{eqnarray}
\Phi_{\rm{tid}}(r)=-\frac{GM}{4r}\left(\frac{a}{r}\right)^2\frac{q}{(1+q)^2}\left(1+\frac{3}{2}e^2\right).
\label{eq:tide}
\end{eqnarray}
From this equation, equation~(\ref{eq:Phi_r}) is rewritten as $\bar{\Phi}(r)=-GM/r
+ \Phi_{\rm{tid}}(r)$.
It is noted from equation~(\ref{eq:tide}) that the tidal force acts more strongly
on the circumbinary disk
around an eccentric binary than around {a} circular binary.  

\subsection{Trapped waves}

Let us now consider oscillations in the form of the normal mode 
which varies as $\exp[i(\omega{t}-k_{r}r-m\phi)]$ with $\omega$ 
being the oscillation frequency and $k_{\rm{r}}$ and $m$ being {the radial and} 
azimuthal wave numbers, respectively.
Then, the local dispersion relation {for these oscillations is given} by
\begin{eqnarray}
(\omega-m\Omega)^2-\kappa^2=k_{r}^2c_{\rm{s}}^2,
\label{eq:ldr}
\end{eqnarray}
where $c_{\rm{s}}$ is the sound speed, 
$\Omega$ is the {rotation} frequency of the circumbinary disk, 
and $\kappa$ is the epicyclic frequency given by
\begin{eqnarray}
\kappa=2\Omega(2\Omega+rd\Omega/dr).
\label{eq:kappa} 
\end{eqnarray}

From equation~(\ref{eq:ldr}), the region where a wave can propagate,
{i.e., $k_{r}^{2}>0$,}
is given by
\begin{eqnarray}
\frac{\omega}{m}<\Omega - \frac{\kappa}{m}
\quad \mbox{or}\quad \frac{\omega}{m}>\Omega + \frac{\kappa}{m}.
\label{eq:pregion}
\end{eqnarray}
In other words,
waves can propagate inside the inner Lindblad resonance (ILR) radius, where
$\omega/m=\Omega-\kappa/m$, or outside the outer Lindblad resonance (OLR) radius,
where $\omega/m=\Omega+\kappa/m$. 
However, we are interested only in the former type of waves with 
$0<\omega/m<\Omega-\kappa/m$, because they are trapped in the region
between the inner radius of the circumbinary disk and the ILR radius, 
and are thus possibly excited. 
All other waves are running waves and will never be excited.

\cite{kato83} pointed out that the only possible global waves 
in nearly Keplerian disks such as circumbinary disks 
are low-frequency, one-armed (i.e., $m=1$) waves. 
Waves with $m \ne 1$ have wavelengths much smaller than $r$, 
and are expected to damp rapidly as a result.
In the case of a circumbinary disk around a binary with nonextreme mass ratio, however, 
mass transfer occurs from two points of the inner edge of 
the circumbinary disk to each component of every binary orbit \citep{haya07}.
This might suggest that two-armed {(i.e., $m=2$)} spiral waves are excited
in the circumbinary disk through coupling with the tidal potential.


\subsection{{Precession frequency and beat frequency}}

From equation (\ref{eq:Phi_r}), the orbital angular frequency 
of the circumbinary disk under the tidal perturbation potential is given by
\begin{eqnarray}
\Omega\simeq\Omega_{\rm{K}}\left[1+\frac{3}{8}\left(\frac{a}{r}\right)^{2}\frac{q}{(1+q)^2}\left(1+\frac{3}{2}e^2\right)\right],
\label{eq:omd}
\end{eqnarray}
where $\Omega_{\rm{K}}=\sqrt{GM/r^{3}}$ 
is the Keplerian angular frequency of the circumbinary disk.
Substituting equation~(\ref{eq:omd}) in equation~(\ref{eq:kappa}), we have
\begin{eqnarray}
\kappa\simeq\Omega_{\rm{K}}\left[1-\frac{3}{8}\left(\frac{a}{r}\right)^{2}\frac{q}{(1+q)^2}\left(1+\frac{3}{2}e^2\right)\right],
\end{eqnarray}
In general, the eigenfrequency and the extent of the propagation region 
can be estimated, using the WKBJ approximation, by searching for $\omega$ that
satisfies $\int k_r dr \sim n\pi$ with an integer $n$, 
where the integration is performed over the propagation region.
Given that the pressure gradient restoring force is much smaller than the gravity in geometrically-thin circumbinary disks and that the waves are trapped in the inner part of the disk,
the precession frequency of an $m$-armed spiral wave, $\omega_{\rm{p},m}$, is approximately given by
\begin{eqnarray}
\omega_{\rm{p},m}=m\Omega-\kappa
\label{eq:op}
\end{eqnarray}
at the inner disk radius, $r=r_{\rm in}$.
More rigorously, in the case of a one-armed ($m=1$) wave, one can show that the precession frequency 
is significantly lower than, but is still of the order of, 
$\Omega(r_{\rm{in}})-\kappa(r_{\rm{in}})$. 
Thus we write it as
\begin{eqnarray}
\frac{\omega_{\rm{p},1}}{\Omega_{\rm{orb}}}
\lesssim
\frac{3}{4}\left(\frac{a}{r_{\rm
in}}\right)^{7/2}\frac{q}{(1+q)^2}\left(1+\frac{3}{2}e^2\right).
\label{eq:op1}
\end{eqnarray}
Note that for $m=1$ modes the precession frequency is positive and 
much lower than the orbital frequency.
This feature means that the $m=1$ waves 
with wavelengths comparable to $r_{\rm in}$ are trapped 
in the inner part of the circumbinary disk.}

On the other hand, the precession frequency 
of an $m$-armed ($m \ne 1$) wave
normalized by the orbital frequency of the binary
is obtained as
\begin{eqnarray}
\frac{\omega_{\rm{p},m}}{\Omega_{\rm{orb}}}
&=&
\left(\frac{a}{r_{\rm in}}\right)^{3/2}
\left[ m-1 \frac{}{} \right. \nonumber \\
&& + \left. \frac{3}{8}(m+1)\left(\frac{a}{r_{\rm in}}\right)^{2}
    \frac{q}{(1+q)^2}\left(1+\frac{3}{2}e^2\right) \right] \nonumber \\
&\simeq& (m-1)\left(\frac{a}{r_{\rm in}}\right)^{3/2}.
\label{eq:op2}
\end{eqnarray}
Thus, the precession frequency of $m \ne 1$ waves is approximately 
given by $(m-1)\Omega_{\rm K}(r_{\rm in})$. Note that these waves have wavelengths
much shorter than those of $m=1$ waves.

It is instructive to evaluate the beat periods for some important modes.
The beat frequency for an $m$-armed spiral wave is defined by
\begin{eqnarray}
\Omega_{\rm{beat},m}=\Omega_{\rm{orb}}-\omega_{\rm{p},m}.
\label{eq:beatf}
\end{eqnarray}
From equation~(\ref{eq:beatf}), we obtain the beat period {as}
\begin{eqnarray}
P_{\rm{beat},m}=P_{\rm{orb}}\left(1-\frac{\omega_{\rm{p},m}}{\Omega_{\rm{orb}}}\right)^{-1}.
\label{eq:beatp}
\end{eqnarray}
According to the tidal truncation theory \citep{al94}, 
the circumbinary disk around the binary with a low-moderate orbital eccentricity 
is truncated at $r_{\rm{in}}\simeq2.08a$.
In an equal-mass binary ($q=1$) with $e=0.5$, 
the beat period of the $m=1$ mode is $P_{\rm{beat},1}\sim1.03P_{\rm{orb}}$,
whereas that of the $m=2$ mode is $P_{\rm{beat},2}\sim1.34P_{\rm{orb}}$.


\section{Summary and Discussion}

We have studied possible oscillation modes 
in a circumbinary disk induced by the tidal potential 
of a binary system such as young binary stars, 
protoplanetary systems, and massive binary black holes in AGNs.
We have pointed out that observationally interesting waves are 
those with precession frequency 
$\omega_{{\rm p},m}/m \lesssim \Omega(r_{\rm in})-\kappa(r_{\rm in})/m$, 
where $r_{\rm in}$ is the inner radius of the circumbinary disk.
These waves are trapped between 
the inner disk radius and the ILR radius,
where the pattern speed of the wave is equal to $\Omega(r)-\kappa(r)/m$.
Among them, only $m=1$ waves have wavelengths comparable 
to the inner disk radius.
{When the $m=1$ mode is dominant, 
the inner region of the circumbinary disk becomes eccentric 
and precesses very slowly \citep[c.f.,][]{kd06,mm08}}.
On the other hand, waves with $m \ne 1$ have much shorter wavelengths and 
affect only the innermost narrow region of the disk.
For example, if the $m=2$ mode is dominant, the disk inner edge is
deformed to an elliptical shape, which precesses at the local Keplerian frequency.

It is important to note that there are two types of excitation mechanism for these waves. 
In circular binaries, nonaxisymmetric perturbations in the disk can grow through the
resonant interaction with the tidal potential at particular resonance radii, 
a mechanism similar to that for superhumps in dwarf novae systems. 
For example, the growth of an eccentric perturbation is driven by the excitation
of an $m=2$ wave at the 1:3 OLR radius.
In addition to this mechanism, in eccentric binaries,
a one-armed ($m=1$) spiral wave can also be excited through direct driving as a result of a one-armed ($m=1$) potential \citep{al96b}.

Whatever the excitation mechanism,
the deformed inner part of the circumbinary disk precesses at
the frequency given by equation~(\ref{eq:op1}) for the $m=1$ mode 
and equation (\ref{eq:op2}) for the $m=2$ mode.
Since the velocity field in the disk is also perturbed by the waves,
the emission line profiles from the inner part of the circumbinary disk
will vary with the precession period.
Such a variability has a distinct feature and will easily be observed.
Since the $m=1$ mode is a low-frequency, eccentric mode, the relative intensity
of the violet (V) and red (R) peaks of double-peaked line profiles
varies with a long period, e.g., $\sim40P_{\rm{orb}}$ for 
an equal-mass binary with $e=0.5$.
Such a line-profile variability caused by $m=1$ waves has long been known 
as the V/R variation in Be stars, 
B-type stars with circumstellar decretion disks \citep[e.g.,][]{por03}.
In contrast, if the $m=2$ mode is dominant,
the double-peaked profiles stay symmetric, 
but their peak separations (and FWHMs) will vary with
a short period of $\sim 2^{3/2} P_{\rm{orb}}$
irrespective of binary parameters.

In addition to these line-profile variabilities,
another type of variability from circumbinary disks is expected.
There is the beat between the orbital frequency of the binary
and the precession frequency of the circumbinary disk.
The beat period is slightly longer than the orbital period 
for $m=1$ and $m=2$ modes, as seen from equations~(\ref{eq:op1}), (\ref{eq:op2}), and (\ref{eq:beatp}).
The radiation emitted from the circumbinary disk through the tidal dissipation 
is expected to vary periodically with the beat period,
because the distance between each component of the binary and 
the inner edge of the circumbinary disk periodically changes.
Moreover, the mass transfer rate from the circumbinary disk to 
each binary component {is also expected to} vary with the same period, 
because of angular momentum removal by the tidal torques.
Therefore, the circumbinary disk perturbed by a trapped density wave
will show the light-curve {modulation} at the beat period.

As for observability, the light-curve modulation caused by an $m=1$ deformation 
is expected to be detected more easily than those caused by the $m\neq1$ one.
This is because the $m=1$ wave can exist globally in the circumbinary disk,
while the $m\neq1$ waves exist only in a narrow region.

In this Letter, we have shown that variabilities with the beat period and/or
the precession period are expected as a natural consequence of 
the tidal interaction between a binary and a circumbinary disk 
and that one can identify circumbinary disks
in terms of these periodic variabilities.
This provides a new method to probe circumbinary disks in various
astrophysical systems. In a forthcoming paper, we will perform
a more detailed analysis of the mode characteristics, including numerical simulations.

\acknowledgments

We thank the anonymous referee for useful comments and suggestions.
K.H. is grateful to Shin Mineshige
for helpful discussions. 
The calculations reported here were 
performed using the facility at the Centre for Astrophysics \& Supercomputing 
at Swinburne University of Technology, Australia and at YITP in Kyoto University. 
This work has been supported in part by 
the Grants-in-Aid of the Ministry of Education, Science, Culture, and 
Sport and Technology (MEXT; 30374218 KH and 20540236 ATO) 
and by the Grant-in-Aid for the 21st Century COE Scientific Research Programs on 
"Topological Science and Technology''  from MEXT.

\end{document}